\documentstyle[seceq,supplement,epsf,wrapft]{ptptex}

\preprintnumber[3cm]{
KUNS-1325\\PTPTeX ver.0.8\\ August, 1997}

\title{
Magnetic Phase Transition and Magnetization Plateau\\
in Cs$_2$CuBr$_4$
}

\author{
Hidekazu {\sc Tanaka}\footnote{e-mail: tanaka@lee.phys.titech.ac.jp}, Toshio {\sc Ono}, Hiroko {\sc Aruga Katori}$^{1}$, 
\\
Hiroyuki {\sc Mitamura}$^{2}$, Fumihiro {\sc Ishikawa}$^{2}$ and Tsuneaki {\sc Goto}$^{2}$
}

\inst{
Department of Physics, Tokyo Institute of Technology, Tokyo 152-8551
\\
$^1$RIKEN (The Institute of Physical and Chemical Research), Wako 351-0198
\\
$^2$Institute for Solid State Physics, The University of Tokyo, Kashiwa 277-8581
}

\recdate{
December 25, 2001}

\abst{
The crystal structure of Cs$_2$CuBr$_4$ is the same as that of Cs$_2$CuCl$_4$, which has been characterized as a spin-$\frac{1}{2}$ quasi-two-dimensional frustrated system. The magnetic properties of Cs$_2$CuBr$_4$ were investigated by magnetization and specific heat measurements. The phase transition at zero magnetic field was detected at $T_{\rm N}=1.4$ K. It was observed that the magnetization curve has a plateau at about one-third of the saturation magnetization for magnetic field $H$ parallel to the $b$- and $c$-axes, while no plateau was observed for $H\parallel a$. The field-induced phase transition to the plateau state appears to be of the first order. The mechanism leading to the magnetization plateau is discussed.
}

\begin{document}

\maketitle


\makeatletter
\if 0\@prtstyle
\def\asp{.3em} \def\bsp{.26em}
\else
\def\asp{.3em} \def\bsp{.3em}
\fi \makeatother

\section{Introduction}

It is known that in the classical calculation, the magnetization curves at zero temperature for triangular antiferromagnets (TAF) with the Heisenberg and $XY$ spins are monotonical up to the saturation. However, recent theoretical studies~\cite{Nishimori}\tocite{Honecker} revealed that the quantum fluctuation produces novel phase transitions accompanied by a plateau with one-third of the saturation magnetization $M_{\rm s}$ or a jump of the magnetization. Such a field-induced quantum phase transition was actually observed in CsCuCl$_3$,~\cite{Motokawa1}\tocite{Schotte} which is described as a ferromagnetically stacked TAF.~\cite{Adachi} The phase transition with a small magnetization jump for the magnetic field $H$ parallel to the $c$-axis arises from the competition between the weak easy-plane anisotropy and the quantum fluctuation, and a new coplanar spin structure, which is unstable in the classical calculation, is realized in the high-field phase.~\cite{Nikuni,Motokawa2,Schotte} However, the experimental realization of the quantum-fluctuation-induced plateau at $M_{\rm s}/3$ has not been reported so far.

Cs$_2$CuBr$_4$ is isomorphous with Cs$_2$CuCl$_4$, which has an orthorhombic structure with space group $Pnma$.~\cite{Helmholtz}\tocite{Li} Figure 1(a) shows the crystal structure of Cs$_2$CuBr$_4$. The structure is composed of CuBr$_4^{2-}$ tetrahedra and Cs$^+$ ions. The tetrahedra are linked along the $b$-axis. The CuBr$_4^{2-}$ tetrahedra are compressed along the axes perpendicular to the $b$-axis because of the Jahn-Teller effect. Figure 1(b) shows the arrangement of the CuBr$_4^{2-}$ tetrahedra in the $bc$-plane. Cu$^{2+}$ ions with spin-$\frac{1}{2}$ form a distorted triangular lattice in the $bc$-plane.
\begin{figure}
            \epsfysize =4 cm
            \centerline{\epsfbox{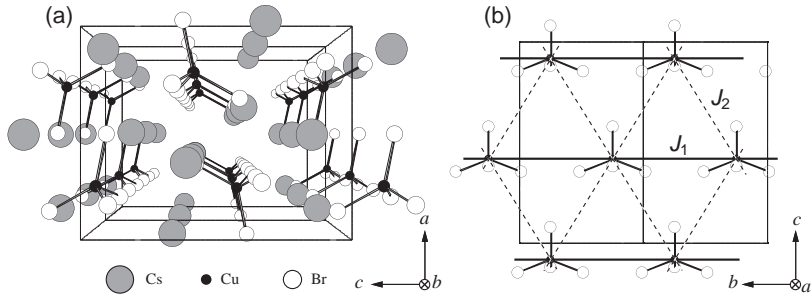}}
        \caption{(a) The perspective view of the crystal structure of Cs$_2$CuBr$_4$ parallel to the $b$-axis. Shaded, closed and open circles denote Cs$^+$, Cu$^{2+}$ and Br$^-$ ions, respectively. (b) The arrangement of the CuBr$_4^{2-}$ octahedra in the $bc$-plane, where Cs$^+$ ions are omitted.}
        \label{fig:1}
\end{figure}

The magnetic properties of Cs$_2$CuCl$_4$ have been extensively investigated by magnetic susceptibility measurement and neutron scattering experiments.~\cite{Carlin}\tocite{Coldea3} The susceptibility of Cs$_2$CuCl$_4$ exhibits a broad maximum at around 3 K,~\cite{Carlin} which was attributed to the one-dimensional (1D) antiferromagnetic exchange coupling along the $b$-axis. 
Cs$_2$CuCl$_4$ undergoes a magnetic phase transition at $T_{\rm N}=0.62$ K.~\cite{Coldea1} Below $T_{\rm N}$, spins form a helical incommensurate structure with an ordering vector $\mib Q=(0, 0.472, 0)$ in a plane which is almost parallel to the $bc$-plane.~\cite{Coldea1} The incommensurate spin structure arises from the condition $J_1\neq J_2$, {\it i.e.,} $\cos (\pi Q)=-J_2/(2J_1)$. The magnetic excitations in Cs$_2$CuCl$_4$ was investigated by means of neutron inelastic scattering, and it was demonstrated that the exchange interactions $J_1$ and $J_2$ are dominant and the interlayer coupling is smaller than $10^{-2}J_1$, {\it i.e.,} Cs$_2$CuCl$_4$ is the quasi-two-dimensional (2D) frustrated spin system.~\cite{Coldea2,Coldea3} Coldea {\it et al.}~\cite{Coldea2,Coldea3} obtained the phase diagram for the magnetic field vs. temperature. They showed that for $H\parallel c$ the ordered phase vanishes before reaching the saturation, and that the spin state between the ordered state and the saturated state is a spin liquid. The present study is motivated by the field-induced vanishment of the ordered phase. However, the ordering temperature $T_{\rm N}=0.62$ K is too low to perform various measurements. Thus, we choose Cs$_2$CuBr$_4$, because the exchange interaction through bromine ions is generally stronger than that through chlorine ions. We have performed magnetization and specific heat measurements to investigate the magnetic phase transitions in Cs$_2$CuBr$_4$. As shown below, unexpected magnetization plateau was observed.

\section{Experimental Procedures}

Single crystals of Cs$_2$CuBr$_4$ were grown by the slow evaporation of aqueous solution of CsBr and CuBr$_2$ in a mole ratio $2:1$. The crystals obtained were identified to be Cs$_2$CuBr$_4$ by X-ray diffraction. The crystal is cleaved along the $(0, 0, 1)$ plane.

The specific heat measurements for single crystal of Cs$_2$CuBr$_4$ were carried out at RIKEN down to 0.6 K using a Mag Lab$^{\rm HC}$ microcalorimeter (Oxford Instruments) in which the relaxation method was employed.
The magnetizations were measured down to 1.8 K in magnetic fields up to 7 T using a SQUID magnetometer (Quantum Design MPMS XL). 
The high-field magnetization measurement was performed using an induction method with a multilayer pulse magnet at the Ultra-High Magnetic Field Laboratory, Institute for Solid State Physics, The University of Tokyo. Magnetization data were collected mainly at $T=0.4$ K in magnetic fields up to 35 T.
\begin{figure}[htb]
\parbox{\halftext}{
	 \epsfysize =5 cm
	 \epsfbox{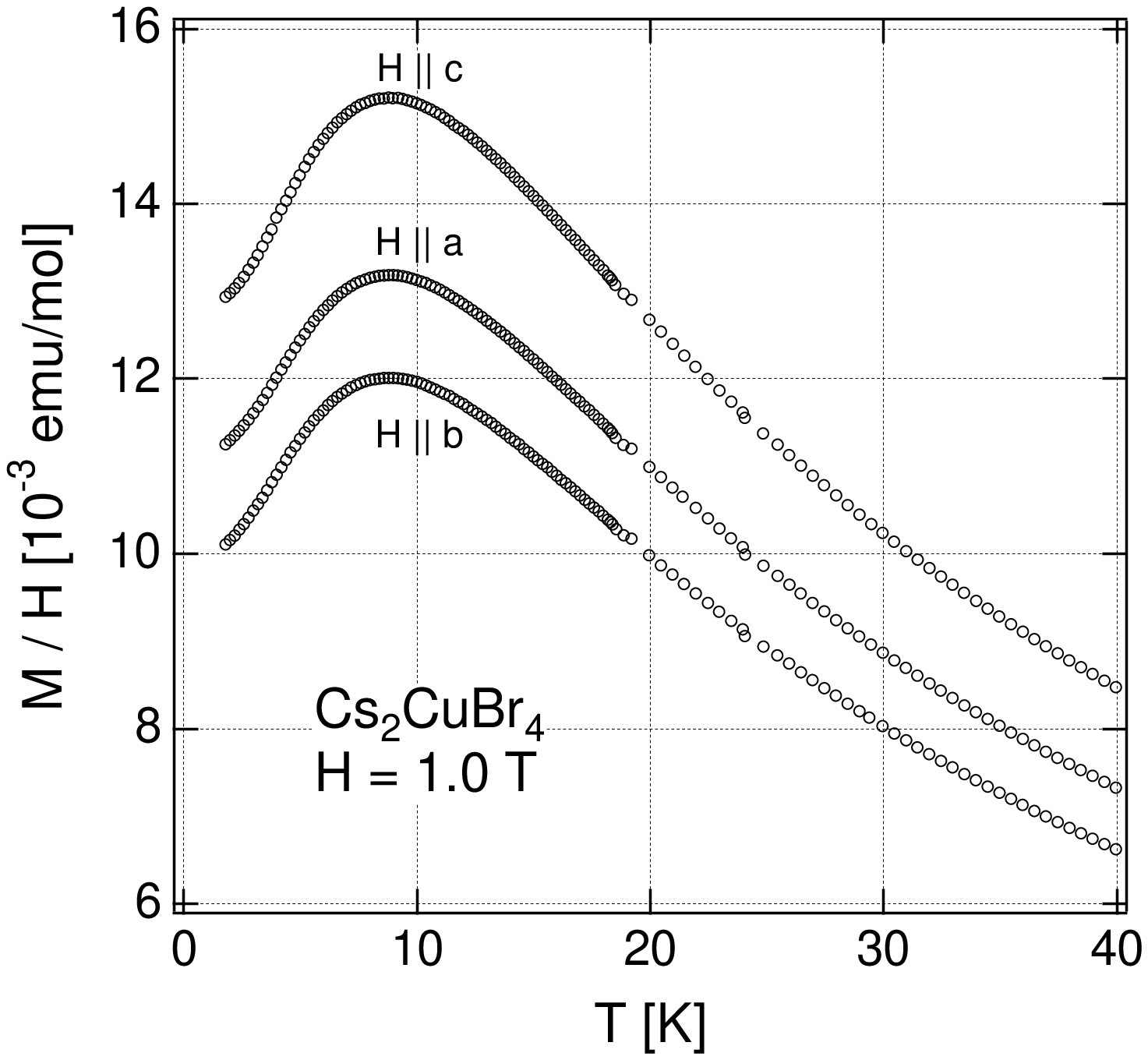}
	\caption{Magnetic susceptibilities in Cs$_2$CuBr$_4$ for $H$ parallel to the $a$-, $b$- and $c$-axes.}
}
\hspace{8mm}
\parbox{\halftext}{
	 \epsfysize =5 cm
	 \epsfbox{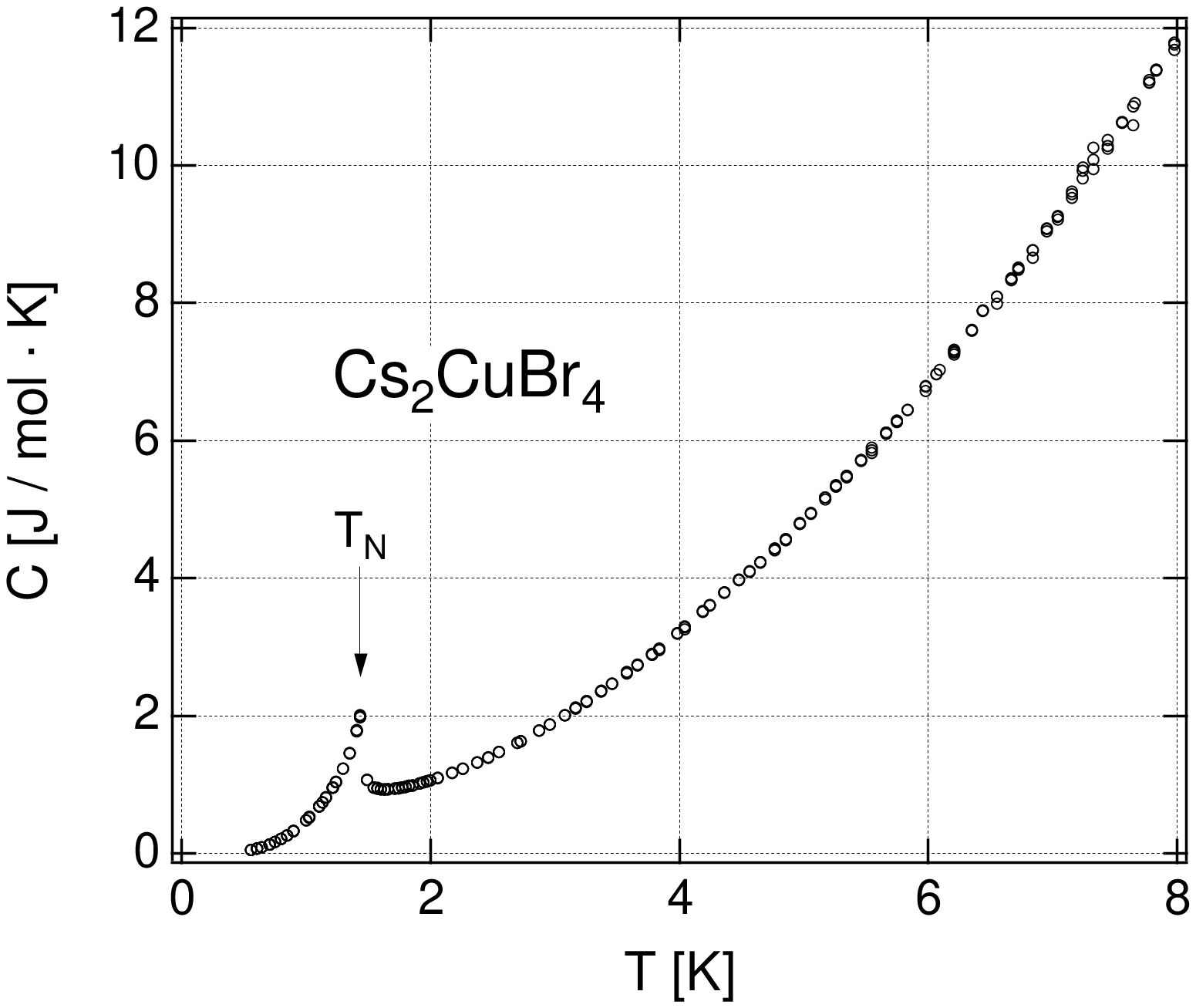}
	\caption{The total specific heat $C$ in Cs$_2$CuBr$_4$ measured at zero field.}
}
\end{figure}

\section{Results and Discussion}

Figure 2 shows the magnetic susceptibilities $M/H$ in Cs$_2$CuBr$_4$ measured at $H=1.0$ T. The susceptibilities for three different field directions exhibit the similar temperature variation. The differences between the absolute values of the three susceptibilities are due to the anisotropy of the $g$-factor. The susceptibilities display broad maxima at $T_{\rm max}\approx 9$ K, which is characteristic of the low-dimensional antiferromagnetic spin system. The value of $T_{\rm max}$ for Cs$_2$CuBr$_4$ is about 3 times as large as $T_{\rm max}\approx 3$\,K for Cs$_2$CuCl$_4$. This implies that the exchange interactions in Cs$_2$CuBr$_4$ are larger than those in Cs$_2$CuCl$_4$ as expected. 

Figure 3 shows the total specific heat $C$ in Cs$_2$CuBr$_4$ at zero field. The $\lambda$-like anomaly indicative of the phase transition is observed at $T_{\rm N}=1.4$ K. The ordering temperature $T_{\rm N}$ for Cs$_2$CuBr$_4$ is more than twice as high as $T_{\rm N}$ K for Cs$_2$CuCl$_4$. 

Figure 4 shows the magnetization curves measured at $T=0.4$ K for $H\parallel a$, $b$ and $c$. The magnetization saturates at $H_{\rm s}\approx 30$ T. The value of the saturation magnetization which is slightly larger than $1 \mu_{\rm B}$ is consistent with spin-$\frac{1}{2}$. This result indicates that the electronic ground state is non-degenerate due to large Jahn-Teller distortion, so that the orbital moment is quenched. Thus, the magnetic properties of Cs$_2$CuBr$_4$ can be described by a spin-$\frac{1}{2}$ Heisenberg model with small anisotropy. The differences between the absolute values of the saturation fields and the saturation magnetizations for three field directions are due to the anisotropy of the $g$-factor. 
\begin{figure}
			\epsfysize = 14 cm
            \centerline{\epsfbox{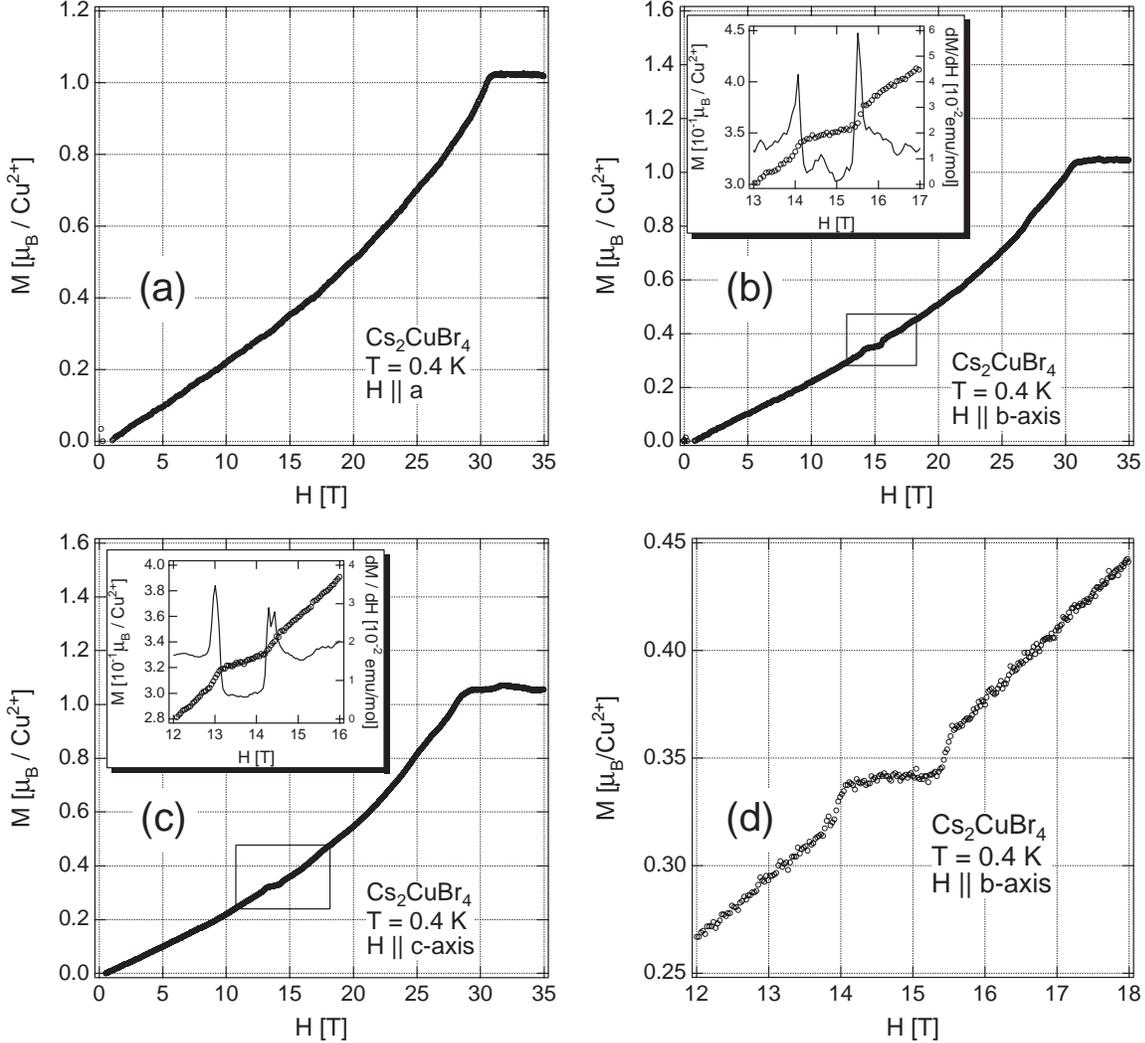}}
        \caption{Magnetization curves in Cs$_2$CuBr$_4$ measured at $T=0.4$ K for (a) $H\parallel a$, (b) $H\parallel b$ and (c) $H\parallel c$. The insets in (b) and (c) show $dM/dH$ vs. $H$ around the plateaus. (d) The magnetization curve for $H\parallel b$ measured in magnetic fields up to 20 T.}
        \label{fig:4}
\end{figure}

The magnetization curve for $H\parallel a$ is monotonical up to the saturation, while the magnetization curves for $H\parallel b$ and $c$ have a plateau at about one-third of the saturation magnetization $M_{\rm s}$. As shown in the insets of Figs. 4(b) and (c), the field derivative of the magnetization $dM/dH$ exhibits sharp peaks at the both edges of the plateau. The level of the plateau is just $M_{\rm s}/3$ for $H\parallel b$, while it is slightly lower than $M_{\rm s}/3$ for $H\parallel c$. The plateau region is narrow, and its field range is about 1.5 T for $H\parallel b$ and 1 T for $H\parallel c$. Figure 4(d) shows the magnetization curve for $H\parallel b$ measured in magnetic fields up to 20 T. Since the highest field is 20 T, the mechanical noise is much suppressed, so that the plateau is clearly observed. The transition between the slope and the plateau regions appears to be discontinuous.
The magnetization plateau was not observed at $T=1.6$ K, which is higher than $T_{\rm N}$. Thus, we can conclude that the plateau exists only in the ordered phase.

The broad maximum of the magnetic susceptibility shown in Fig. 2 is suggestive of a spin-$\frac{1}{2}$ antiferromagnetic Heisenberg chain.
If this is the case, then the saturation field $H_{\rm s}$ is calculated as $H_{\rm s}=3.12k_{\rm B}T_{\rm max}/g\mu_{\rm B}\approx 20$ T with $T_{\rm max}\approx 9$ K.~\cite{Bonner} This value is two-third of the observed value $H_{\rm s}\approx 30$ T. This implies that the interchain interaction ($J_2$) is significantly large as observed in Cs$_2$CuCl$_4$.

At the moment, the spin structure below $T_{\rm N}$ and the details of the exchange interactions in Cs$_2$CuBr$_4$ are not known. However, since the isostructural Cs$_2$CuCl$_4$ has a two-dimensional exchange network with the frustration and a helical spin structure in the $bc$-plane,~\cite{Coldea3} we assume that Cs$_2$CuBr$_4$ also has the similar properties. 

The classical molecular field theory~\cite{Nagamiya1} predicts that a transition from a helical spin structure to a fan structure occurs, when an external field is applied in the easy-plane. An example of the helix-fan transition was recently found in RbCuCl$_3$.~\cite{Maruyama} The helix-fan transition is accompanied with the jump of magnetization, and not with the plateau. 

In the classical TAF with the easy-axis anisotropy, a magnetization plateau can exist at $M_{\rm s}/3$, when an external field is applied along the easy-axis.~\cite{Miyashita} The plateaus due to this classical mechanism have been observed in RbFe(MoO$_4$)$_2$ and CsFe(SO$_4$)$_2$,~\cite{Inami} and GdPd$_2$Al$_3$.~\cite{Kitazawa} However, the magnetization curve has no plateau, when an external field is perpendicular to the easy-axis. 

In the classical Heisenberg TAF, a collinear up-up-down ({\it uud}) spin structure can be stabilized by the thermal fluctuation.~\cite{Kawamura} Consequently, the magnetization curve can have a plateau at $M_{\rm s}/3$, although the plateau is smeared due to the finite temperature effect. With decreasing temperature, the field range of the plateau decreases and vanishes at $T=0$, because the thermal fluctuation is reduced.

Since the magnetization plateau in Cs$_2$CuBr$_4$ is clearly observed for two different field directions and at $T=0.4$ K, which is much lower than $T_{\rm N}$, the plateau cannot be interpreted in terms of the classical model. Thus, the magnetization plateau should be attributed to the quantum effect. In the spin-$\frac{1}{2}$ Heisenberg TAF, the quantum fluctuation removes the continuous degeneracy of the ground state spin configuration in the magnetic field, which cannot be removed by the classical approach, and stabilizes the {\it uud} spin structure in the finite field range to produce a plateau at $M_{\rm s}/3$.~\cite{Nishimori}\tocite{Honecker} In the spin wave approach, the quantum fluctuation is represented by the zero-point oscillation.~\cite{Chubukov}\tocite{Jacobs1} Since the magnetic properties of Cs$_2$CuBr$_4$ may be described by the distorted TAF in the $bc$-plane as observed in Cs$_2$CuCl$_4$, we infer that the magnetization plateau is produced by the mechanism similar to that for the spin-$\frac{1}{2}$ Heisenberg TAF. Therefore, we suggest that in the present system, the {\it uud} spin structure or closely related structure is realized at the plateau due to the quantum fluctuation.

As previously mentioned, the transition to the plateau state appears to be of the first order. This suggests that the ordering vector $\mib Q$ varies discontinuously at the transition field. Thus, it is considered that the spin structures outside the plateau state is incommensurate as observed in Cs$_2$CuCl$_4$,~\cite{Coldea3} while in the plateau state, the spin structure is locked into the collinear {\it uud} structure. 

Since Cs$_2$CuBr$_4$ has the same crystal structure as Cs$_2$CuCl$_4$ has, it is expected that Cs$_2$CuBr$_4$ has the easy-plane anisotropy, which constrains spins to lie in the $bc$-plane as observed in Cs$_2$CuCl$_4$. The anisotropy prevents the realization of the collinear {\it uud} structure for $H\parallel a$.~\cite{Nikuni} Thus, the absence of the magnetization plateau for $H\parallel a$ should be ascribed to the easy-plane anisotropy.

\section{Conclusions}

We have presented the results of magnetization and specific heat measurements on the quantum spin system Cs$_2$CuBr$_4$, which may be described by the spin-$\frac{1}{2}$ Heisenberg antiferromagnet on the distorted triangular lattice with the small easy-plane anisotropy. The magnetization plateau was observed at about one-third of the saturation magnetization for $H\parallel b$ and $H\parallel c$, while no plateau was observed for $H\parallel a$. It was suggested that the plateau is produced by the interplay of the spin frustration and the quantum fluctuation. The transition to the plateau state appears to be of the first order, which is indicative of the incommensurate-commensurate transition.

\section*{Acknowledgements}
The authors would like to thank T. Sakai and S. Miyashita for useful comments on the magnetization plateau. This work was supported by Toray Science Foundation and a Grant-in-Aid for Scientific Research on Priority Areas (B) from the Ministry of Education, Culture, Sports, Science and Technology of Japan.

\end{document}